\documentclass[%
 aip,
 amsmath,amssymb,
 reprint,%
]{revtex4-1}
\usepackage{graphicx}
\usepackage{dcolumn}
\usepackage{bm}

\usepackage[utf8]{inputenc}
\usepackage[T1]{fontenc}
\usepackage{mathptmx}
\usepackage{etoolbox}
\usepackage{xcolor}
\usepackage{siunitx}

\makeatletter
\def\@email#1#2{%
 \endgroup
 \patchcmd{\titleblock@produce}
  {\frontmatter@RRAPformat}
  {\frontmatter@RRAPformat{\produce@RRAP{*#1\href{mailto:#2}{#2}}}\frontmatter@RRAPformat}
  {}{}
}%
\makeatother

\draft 

\begin{document}

\preprint{AIP/123-QED}

\title[Radiation and Heat Transport in Divergent Shock-Bubble Interactions]{Radiation and Heat Transport in Divergent Shock-Bubble Interactions} 



\author{K. Kurzer-Ogul}
\email{kkurzero@ur.rochester.edu}
\affiliation{Department of Mechanical Engineering, University of Rochester, Rochester, New York 14620, USA}
\affiliation{Los Alamos National Laboratory, Los Alamos, New Mexico 87545, USA}
\author{B. M. Haines}
\author{D. S. Montgomery}
\affiliation{Los Alamos National Laboratory, Los Alamos, New Mexico 87545, USA}
\author{S. Pandolfi}
\affiliation{SLAC National Accelerator Laboratory, Menlo Park, California 94025, USA}
\author{J. P. Sauppe}
\affiliation{Los Alamos National Laboratory, Los Alamos, New Mexico 87545, USA}
\author{A. F. T. Leong}
\affiliation{Los Alamos National Laboratory, Los Alamos, New Mexico 87545, USA}
\author{D. Hodge}
\affiliation{Department of Physics and Astronomy, Brigham Young University, Provo, Utah 84602, USA}
\author{P. M. Kozlowski}
\affiliation{Los Alamos National Laboratory, Los Alamos, New Mexico 87545, USA}
\author{S. Marchesini}
\author{E. Cunningham}
\author{E. Galtier}
\author{D. Khaghani}
\author{H. J. Lee}
\author{B. Nagler}
\affiliation{SLAC National Accelerator Laboratory, Menlo Park, California 94025, USA}
\author{R. L. Sandberg}
\affiliation{Department of Physics and Astronomy, Brigham Young University, Provo, Utah 84602, USA}
\author{A. E. Gleason}
\email{ariannag@stanford.edu}
\affiliation{SLAC National Accelerator Laboratory, Menlo Park, California 94025, USA}
\author{H. Aluie}
\author{J. K. Shang}
\affiliation{Department of Mechanical Engineering, University of Rochester, Rochester, New York 14620, USA}
\affiliation{Laboratory for Laser Energetics, University of Rochester, Rochester, New York 14620, USA}
\date{\today}

\begin{abstract}
\begin{quotation}
The following article has been accepted by {\em Physics of Plasmas}. After it is published, it will be found at \url{https://doi.org/10.1063/5.0185056}.
\end{quotation}
Shock-bubble interactions (SBI) are important across a wide range of physical systems. In inertial confinement fusion, interactions between laser-driven shocks and micro-voids in both ablators and foam targets generate instabilities that are a major obstacle in achieving ignition. Experiments imaging the collapse of such voids at high energy densities (HED) are constrained by spatial and temporal resolution, making simulations a vital tool in understanding these systems. In this study, we benchmark several radiation and thermal transport models in the xRAGE hydrodynamic code against experimental images of a collapsing mesoscale void during the passage of a 300 GPa shock. We also quantitatively examine the role of transport physics in the evolution of the SBI. This allows us to understand the dynamics of the interaction at timescales shorter than experimental imaging framerates. We find that all radiation models examined reproduce empirical shock velocities within experimental error. Radiation transport is found to reduce shock pressures by providing an additional energy pathway in the ablation region, but this effect is small ($\sim$1\% of total shock pressure). Employing a flux-limited Spitzer model for heat conduction, we find that flux limiters between 0.03 and 0.10 produce agreement with experimental velocities, suggesting that the system is well-within the Spitzer regime. Higher heat conduction is found to lower temperatures in the ablated plasma and to prevent secondary shocks at the ablation front, resulting in weaker primary shocks. Finally, we confirm that the SBI-driven instabilities observed in the HED regime are baroclinically driven, as in the low energy case.
\end{abstract}

\maketitle 

\section{\label{sec:level1}Introduction}
Voids are low-density regions in an otherwise higher-density medium. Because of the ubiquity of voids in all physical media, the hydrodynamics of their collapse are important across a wide range of physics. The collapse of voids in response to an impulsive shock, a type of divergent shock-bubble interaction (SBI), appears in areas such as fusion energy science\cite{zylstraHotspotMixLargescale2020,hainesMechanismReducedCompression2022}, supersonic combustion\cite{yangModelCharacterizationVortex1994}, foam-based shock mitigation systems\cite{ballShockBlastAttenuation1999, delaleDirectNumericalSimulations2005}, shock wave lithotripsy\cite{deliusExtracorporealShockWaves1998, eisenmengerMechanismsStoneFragmentation2001}, atmospheric science\cite{davyMeasurementsRefractionDiffraction2005}, and astrophysical flows\cite{kleinHydrodynamicInteractionShock1994,arnettRoleMixingAstrophysics2000}. SBIs are especially important in astrophysics, where they affect the dynamics of supernovae\cite{gamezoThermonuclearSupernovaeSimulations2003} and the interaction of their remnants with the interstellar medium (ISM)\cite{kleinInteractionSupernovaRemnants2003, hansenExperimentMassstrippingInterstellar2007}. Strong shocks originating from supernovae, stellar winds, expanding nebulae and spiral density waves are common in interstellar space, where they severely disrupt atomic and molecular clouds\cite{kleinHydrodynamicInteractionShock1994, ranjan_shock-bubble_2011}. This results in turbulent mixing and momentum and energy exchange between the cloud and ISM gases\cite{ranjan_shock-bubble_2011}, which in turn play a vital role in star formation and the evolution of galaxies\cite{hansenExperimentMassstrippingInterstellar2007}. These shock-cloud interactions have been studied in scaled laboratory experiments using SBI configurations\cite{kleinHydrodynamicInteractionShock1994,kleinInteractionSupernovaRemnants2003,hansenExperimentMassstrippingInterstellar2007}.

In the high energy density (HED) regime, instabilities generated by SBIs are a major challenge in the pursuit of fusion as a viable clean energy source. In inertial confinement fusion (ICF), a fuel pellet is compressed through the application of laser energy which ablates the outer layer of the capsule, launching a shock inward; it is well-established that Rayleigh-Taylor and Richtmyer-Meshkov instabilities at the fuel-ablator interface significantly degrade the performance of the fuel capsule \cite{betti2016inertial,woo2018effects}, and that these instabilities arise largely due to mesoscale defects (voids) in the ablator material introduced during manufacturing\cite{yan2016three,zhang2018nonlinear}. Rayleigh-Taylor growth of density perturbations associated with voids can puncture the ablator\cite{zhang2020nonlinear} and thereby seed asymmetric compression and jetting of material into the fuel region\cite{woo2018impact}. Despite the recent successes at the National Ignition Facility (NIF)\cite{zylstraBurningPlasmaAchieved2022,zylstraExperimentalAchievementSignatures2022}, several orders of magnitude of efficiency gain are required for any ICF-based power generation scheme. Therefore, addressing materials challenges, including instability mitigation and/or control, are still crucial for success, and it is vital to understand the development of instabilities arising from the shock-induced collapse of these voids.

Similarly to the related problem of shock-induced particle deformation\cite{zhang2003shock,ripley2007acceleration,ling2013shock,acharya2022numerical}, shock-induced void collapse involves highly non-linear processes, making it a difficult problem to study. These include shock reflection, refraction and focusing, which result in the formation of plasma jets and phase transformation in the surrounding material. Several experimental methods have been employed to directly capture images of a void at different stages of collapse. Haas and Sturtevant\cite{haasInteractionWeakShock1987} first used basic optical shadowgraphy to observe the collapse of low-density soap bubbles in a nitrogen-filled shock tube. Ranjan et al.\cite{ranjanShockbubbleInteractionsFeatures2008} improved upon this setup with a free-falling bubble in various background gases and employed laser-based planar Mie scattering diagnostics to observe a 2D cross-section of the flow field; Haehn et al.\cite{haehn_experimental_2011} augmented these techniques with a particle image velocimetry analysis. Both dynamically- and shock-compressed voids have been imaged during collapse in solid silicon and TNT using pulsed X-ray imaging\cite{armstrongTimeresolvedXrayImaging2018,armstrongTimeresolvedXrayImaging2019}. Recently, X-ray phase contrast imaging (XPCI) has been used to image the collapse of mesoscale voids in response to laser-driven shocks at HED pressures\cite{hodgeMultiframeUltrafastXray2022}. Larger scale HED SBI experiments have also been studied with computer vision techniques \cite{kozlowskiUseComputerVision2021}. However, experiments are limited in their ability to elucidate the fine details of the collapse process, because they are subject to physical constraints on spatial and temporal resolution. 

In this study, we use the radiation-hydrodynamic code xRAGE \cite{gittingsRAGERadiationhydrodynamicCode2008,hainesHighresolutionModelingIndirectly2017,hainesDevelopmentHighresolutionEulerian2022} to understand the evolution of a collapsing void by revealing dynamics at timescales shorter than experimental imaging framerates and at sub-experimental resolutions. xRAGE is chosen due to its Eulerian hydrodynamics solver and adaptive mesh refinement (AMR) capability, which makes it uniquely well-suited to modeling the multi-scale physics of a collapsing void. We perform 2D simulations of the experimental platform described in Hodge et al.\cite{hodgeMultiframeUltrafastXray2022} and Pandolfi et al.\cite{pandolfiNovelFabricationTools2022}, wherein a hollow glass sphere embedded in a solid target is subjected to laser-driven shock pressures of over 300 GPa. These simulations are performed with various radiation and heat transport options and are then benchmarked against experimental XPCI images to determine which parameter choices are most appropriate, as well as how radiation and heat conduction affect the development of the system. These experimental images were obtained at the Matter in Extreme Conditions (MEC) instrument\cite{naglerMEC2015,glenzerMEC2016,naglerPCIInstrument2016} at the Linac Coherent Light Source (LCLS)\cite{hodgeMultiframeUltrafastXray2022,pandolfiNovelFabricationTools2022}. Finally, the simulated density and pressure fields are examined to determine the mechanism behind the development of hydrodynamic instabilities.

The rest of this paper is laid out as follows. In Sec. II, we describe the experimental platform being modeled. In Sec. III, we provide background on the xRAGE code and the physics options used in our simulations. In Sec. IV, we discuss the effect of various transport parameters on simulated outcomes and compare our results with experimental benchmarks. We discuss the development of hydrodynamic instabilities in Sec. V. Our conclusions are presented in Sec. VI.

\section{Experimental Platform}
\begin{figure}
\includegraphics[width=0.5\textwidth]{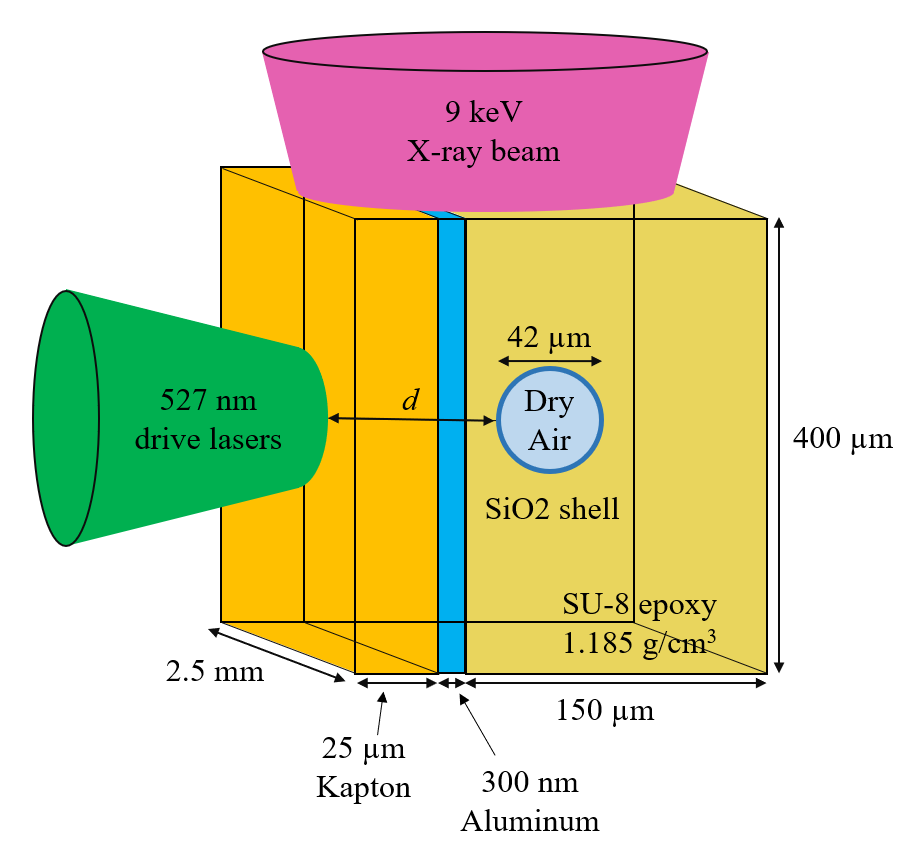}
\caption{\label{fig:fig_1} A schematic of the simulated experiment showing the geometry of the target and laser. The drive laser comprised two beams incident on the Kapton ablator surface from the left. These beams were oriented at 20 degrees from the normal vector to the ablator surface and employed a 150 µm diameter phase plate to irradiate a circular region with a 98 µm Gaussian radius. A secondary beam oriented at 90 degrees to the drive beams was used for X-ray imaging. The glass shell was located at varying distances $d$ from the drive surface in different targets. The target referenced for benchmarking in this work had $d$ = 80 µm.}
\end{figure}

We simulate the experiment described in Hodge et al.\cite{hodgeMultiframeUltrafastXray2022} and Pandolfi et al.\cite{pandolfiNovelFabricationTools2022}, which studies the shock-induced collapse of a 42 µm diameter hollow $\mathrm{SiO}_2$ glass sphere. A schematic of the experimental setup is shown in Fig.~\ref{fig:fig_1}%
. 
A target package containing the sphere is irradiated by a drive laser ($\lambda = 527\ \mathrm{nm}$), which launches an ablative shock into the target. The drive beam uses a 150 µm diameter phase plate and has a Gaussian radius of 98 µm. The sphere consists of a 1 µm-thick glass shell, simulated here with a density of $\rho = 2.20\ \mathrm{g}/\mathrm{cm}^3$, surrounding a ~40 µm diameter interior void composed of dry air at ambient pressure. The glass shell is embedded in a homogeneous block of SU-8 photoresist material\cite{pandolfiNovelFabricationTools2022,shawNegativePhotoresistsOptical1997}, $\,\rho = 1.185\pm0.014\ \mathrm{g}/\mathrm{cm}^3$, which we simulate using a polyamide-imide (PAI) equation of state. SU-8 is chosen for the bulk target due to its similarity to ICF ablator materials\cite{hodgeMultiframeUltrafastXray2022}, particularly in its shock-response, and due to target fabrication considerations\cite{pandolfiNovelFabricationTools2022}. The drive laser produces a 10 ns square pulse which delivers 76.2 J to a target comprising a 25 µm black Kapton CB ablator ($\rho = 1.42\ \mathrm{g}/\mathrm{cm}^3$), a 300 nm aluminum reflective layer, and the void-containing SU-8 block. These material parameters are summarized in Table~\ref{tab:table1}.
\begin{table}
\caption{\label{tab:table1}Density $\rho$ and thickness $d$ of associated layer in the simulated target for each material. The SESAME\cite{lyonSesameAlamosNational,mchardyIntroductionTheoryUse2018} equation of state and opacity used for each material is included as well. Dry air is also used as the background gas.}
\begin{ruledtabular}
\begin{tabular}{ccrcc}
\mbox{Material} & $\rho$ & \multicolumn{1}{c}{$d$} & \mbox{EOS} & \mbox{Opacity}\\
 & $\mathrm{g/{cm}^3}$ & \multicolumn{1}{c}{$\mu\mathrm{m}$} & & \\
\hline
\mbox{Kapton}\footnote{Uses LEOS\cite{moreNewQuotidianEquation1988,youngNewGlobalEquation1995} equation of state}&1.42&25.0&$\mathrm{C}_{22}\mathrm{H}_{10}\mathrm{N}_{2}\mathrm{O}_{5}$&$\mathrm{C}_{22}\mathrm{H}_{10}\mathrm{N}_{2}\mathrm{O}_{5}$\\
Al & 2.70 & 0.3 & Al & Al \\
SU-8 & $1.185$ & 200.0 & $\mathrm{C}_{22}\mathrm{H}_{14}\mathrm{N}_{2}\mathrm{O}_{3}$ & $\mathrm{C}_{22}\mathrm{H}_{14}\mathrm{N}_{2}\mathrm{O}_{3}$\\
Glass & 2.20 & 1.0 & $\mathrm{SiO}_{2}$ & $\mathrm{SiO}_{2}$\\
Dry Air & 1.225e-3 & 40.0 & Dry Air & Dry Air
\end{tabular}
\end{ruledtabular}
\end{table}

The physical target was imaged with a separate X-ray beam in two distinct sets of experiments conducted with the MEC instrument at the LCLS X-ray free electron laser (XFEL)\cite{glenzerMEC2016,naglerMEC2015,naglerPCIInstrument2016}. Single-shot, high-resolution XPCI imaging was performed at an X-ray energy of 17 keV. Multi-frame imaging was performed at 9 keV photon energy using an ultrafast X-ray imaging (UXI) detector\cite{hodgeMultiframeUltrafastXray2022}; in this case, phase contrast was minimized to facilitate image interpretation due to lower spatial resolution. For multi-frame imaging, the XFEL beam was temporally modified to produce four X-ray pulses, allowing the acquisition of multiple images from a single sample and minimizing the uncertainty due to shot-to-shot variations. These images, as well as derived shock speed and pressure measurements, are used to benchmark our simulations in Sec. IV.

\section{The xRAGE Code}
xRAGE \cite{gittingsRAGERadiationhydrodynamicCode2008,hainesHighresolutionModelingIndirectly2017,hainesDevelopmentHighresolutionEulerian2022} is an Eulerian multi-material radiation-hydrodynamic code developed at Los Alamos National Laboratory. It uses AMR (discussed below) and supports various geometries in up to three dimensions. xRAGE is a multiphysics code with packages for radiation and heat transport, laser physics, three temperature plasma physics, various equation of state models, material strength, gravity, turbulence modeling, thermonuclear burn and charged particle transport, and high explosives\cite{hainesDevelopmentHighresolutionEulerian2022}. Because of these capabilities, xRAGE is commonly used in computational studies of ICF implosions and high energy astrophysics, as well as in modeling of HED physics experiments in ICF and laboratory astrophysics, all of which regularly involve SBIs. It has also been used extensively to model shock tube experiments such as those that image SBIs at room temperature.

xRAGE employs a finite volume Godunov-like scheme for the core hydrodynamic solver reminiscent of the method of Harten-Lax-van Leer \cite{hartenUpstreamDifferencingGodunovType1983,vanleerUltimateConservativeDifference1997}. It solves the compressible Euler equations for mass, momentum, and energy conservation:
\begin{eqnarray}
\frac{\partial\rho\mu_k}{\partial t} + \frac{\partial\rho\mu_k u_j}{\partial x_j} = 0,
\\
\frac{\partial\rho u_i}{\partial t} + \frac{\partial\rho u_i u_j}{\partial x_j} + \frac{\partial P}{\partial x_i} = 0,
\\
\frac{\partial\rho\left[\frac{1}{2}u^2 + e \right]}{\partial t} + \frac{\partial\rho u_j\left[\frac{1}{2}u^2 + e \right]}{\partial x_j} + \frac{\partial P u_j}{\partial x_j} = 0,
\label{eq:one}
\end{eqnarray}
along with the pressure-temperature equilibrium condition:
\begin{equation}
\frac{1}{\rho} = \sum^N_{k=1}\mu_k e_k\left(P, T\right).    
\end{equation}
Here, $\rho$ denotes mass density, $u$ is the fluid velocity, $\mu_k$ is the mass fraction for species $k$, $P$ is the pressure, and $e_k$ is the specific internal energy. Together with material equations of state $e_k(P,T)$, this system is solved for pressure and temperature given the total density, specific internal energy, and mass fractions of the component materials\cite{groveXRageHydrodynamicSolver2019}. Godunov methods are variations of the finite volume scheme proposed by Godunov\cite{godunovDifferenceMethodNumerical1959} to solve one-dimensional initial value problems for hyperbolic systems with scale invariant initial data (Riemann problems). xRage has both directionally split\cite{strangConstructionComparisonDifference1968,gittingsRAGERadiationhydrodynamicCode2008} and un-split hydrodynamic solver options that use higher order Godunov schemes. The simulations in this work employ directional splitting.

Both two- and three-temperature plasma physics are available in xRAGE; in the present study we use the latter. A two-temperature (2T) plasma model treats plasma as has having a distinct material and radiation temperature, and a 3T model further splits the material temperature into electron and ion components. This treatment is valid when the timescales of electron-electron and of ion-ion thermal equilibration are much shorter than that of the equilibration between electrons and ions, as in the present case. In addition to the Euler equations and a radiation diffusion equation, the 3T plasma model in xRAGE solves for electron and ion internal energy by considering electron-ion and electron-radiation coupling. This process is described in detail in Refs. \onlinecite{gittingsRAGERadiationhydrodynamicCode2008,hainesHighresolutionModelingIndirectly2017,hainesDevelopmentHighresolutionEulerian2022}.
 
xRAGE utilizes AMR, a mesh adaptation methodology that allows simulation resolution to be changed throughout the computational domain during the course of a simulation in order to balance the need to resolve small features in regions with complex geometries or complex flows and the need to minimize computational expense. xRAGE implements continuous AMR on a grid divided into square cells. The initial refinement level, i.e., cell size, is defined by the user. Further refinement for each cell is then determined according to user-specified criteria based on density and pressure derivatives, the presence of a material interface, and various other considerations\cite{gittingsRAGERadiationhydrodynamicCode2008,hainesHighresolutionModelingIndirectly2017,hainesDevelopmentHighresolutionEulerian2022}. For each refinement level, a cell is subdivided into $2^n$ new cells, where $n$ is the number of dimensions in the problem. Because the refinement is spatially continuous, the size ratio between any two neighboring cells is limited to 1:1 or $2^{n-1}$:1. Refinement continues up to a maximum refinement level set by the user (100 nm in this work) and repeats after every time step. In this way, areas of interest are continuously modeled at high resolution while other areas of the solution space are modeled at reduced resolution, allowing physics of interest to be modeled in higher detail than would otherwise be possible at any given availability of computational resources.

Laser physics are handled in xRAGE through a variant of the Mazinisin laser ray-tracing code from the Laboratory for Laser Energetics \cite{marozasFirstObservationCrossBeam2018,marozasWavelengthdetuningCrossbeamEnergy2018,hainesCouplingLaserPhysics2020}. Laser beams are discretized as rays whose paths are dictated by 3D geometric optics. Energy deposition by the rays is modeled as inverse bremsstrahlung radiation, accounting for the Langdon effect\cite{langdonNonlinearInverseBremsstrahlung1980} and cross-beam energy transfer\cite{randallTheorySimulationStimulated1981}. The ray-trace uses a separate mesh from the radiation-hydrodynamic calculations due to differing refinement needs. 

Equations of state in xRAGE are modeled either analytically, or, more commonly, from pressure-temperature tables. In the present work we use tabular equation of state values from the SESAME equation of state library\cite{lyonSesameAlamosNational,mchardyIntroductionTheoryUse2018}, or in the case of Kapton, from the LEOS library maintained by Lawrence Livermore National Laboratory \cite{moreNewQuotidianEquation1988,youngNewGlobalEquation1995}.

There are several radiation and heat transport \cite{voldMultispeciesPlasmaTransport2019} models available in xRAGE that are of interest to this work. These are described in the following subsections.

\subsection{\label{sec:level2}Radiation}
Multiple radiation solvers are available in xRAGE. The default option is a gray diffusion approximation to solving the full, angular-dependent radiation transport equation\cite{pomraningEquationsRadiationHydrodynamics1973}
\begin{equation}
\frac{1}{c}\frac{\partial I}{\partial t} + \Omega\cdot\nabla I + \sigma I = \frac{c}{4\pi}\left(\sigma_a E_b + \sigma_s E_{r} \right),
\end{equation}
where the intensity $I$ is a function of position, time, and solid angle $\Omega$; $\sigma$, the opacity, is the sum of the absorption and scattering coefficients $\sigma_a$ and $\sigma_s$; and $E_{b,r}$ are the blackbody and radiation energy densities respectively. Eqn. (5) is a gray radiation equation, meaning that it represents the total radiation transport and does not include frequency as a parameter. It uses a single opacity calculated as a frequency-average weighted by a Planckian distribution. This equation can be integrated over $\Omega$ to explicitly include the radiative flux $\boldsymbol{F}$\cite{levermoreFluxlimitedDiffusionTheory1981}:
\begin{equation}
\frac{\partial E}{\partial t} + \nabla\cdot\boldsymbol{F} = c\sigma_a \left(E_b - E_{r} \right).
\end{equation}
In the limit of transport through optically thick (scattering-dominant) material, one can make the diffusion approximation, in which $\boldsymbol{F}$, as derived by Levermore and Pomraning\cite{levermoreFluxlimitedDiffusionTheory1981}, is approximated as
\begin{equation}
\boldsymbol{F} = -\frac{c}{\sigma}D_F\nabla E_r,
\end{equation}
where $D_F$ is a flux-limited diffusion coefficient. The diffusive approximation supports two temperature non-local thermal equilibrium physics (not to be confused 2T or 3T plasma physics, the latter of which this model is a component). There is an additional multi-group option in which radiation energy is divided into frequency groups \cite{winslowMultifrequencyGrayMethodRadiation1995}. Multi-group formulations of Eqns. (6) and (7) are solved for each group independently using the temperature at the beginning of the time step, and the resulting energy spectrum is used to solve the full gray system for the material internal energy coupling.

In the case that multi-group diffusion is insufficient, the Cassio version of xRAGE contains an implementation of the Jayenne implicit Monte Carlo (IMC) solver\cite{thompson21} and the deterministic Capsaicin (Sn) discrete coordinates solver\cite{thompson05} as alternative radiation transport models. In our simulations, we found flux-limited single- and multi-group gray diffusion to provide good agreement with experiment, with marginal (and experimentally unresolvable) deviations from the gray IMC model.

\subsection{\label{sec:level3}Heat Conduction}
 xRAGE features models for both electron and ion thermal conduction. With 3T plasma physics enabled, these solve the equation:
\begin{equation}
\rho C_{v,e,i}\frac{\partial T_{e,i}}{\partial t}=\nabla\cdot\left[\kappa\left(\rho, T_{e,i}\right)\nabla T_{e,i}\right]
\end{equation}
for electron and ion temperatures, respectively. Here $\rho$ is density, $C_v$ is the specific heat capacity for the particle species ($e$ or $i$), $\kappa$ is the normal heat conduction coefficient, and $T$ is the temperature of the particle species. These models treat the energy flux anywhere in space as proportional to the gradient of temperature in the manner of Spitzer and H{\"a}rm\cite{spitzerTransportPhenomenaCompletely1953}. Several conductive models using this approach are available in xRAGE for both electron and ion conduction. Our simulations use a generalized Spitzer conduction coefficient for electron conduction, which in CGS units are:
\begin{equation}
    \kappa = \alpha_k\left(\frac{8}{\pi}\right)^{3/2}\frac{k_B^{7/2}}{e^4\sqrt{m_e}}\frac{T_e^{5/2}}{\ln{\Lambda_{ei}}},
\end{equation}
where $k_B$ is the Boltzmann constant, $e$, $m_e$ and $T_e$ are respectively the electron charge, mass and temperature, and $\ln{\Lambda_{ei}}$ is the Coulomb logarithm. $\alpha_k$ are given in Ref.~\onlinecite{molvigClassicalTransportEquations2014}. The conduction coefficient for ions is
\begin{equation}
    \kappa = 20\left(\frac{2}{\pi}\right)^{3/2}\frac{k_B^{7/2}}{e^4\sqrt{m_i}}\frac{T_i^{5/2}}{{\bar{Z}}^4\ln{\Lambda_{ei}}}\delta_i ,\quad \delta_i = 0.4584,
\end{equation}
as derived by Hayes\cite{hayesElectronIonConductivity2016} based on the model of Lee and More\cite{leeElectronConductivityModel1984}.

Spitzer-H{\"a}rm algorithms are linearized perturbative approximations. These approximations are valid when temperature gradients are small, but become unphysical when the mean free path of thermal electrons exceeds the temperature gradient scale length $T_{e}/\nabla T_{e}$, such as in the vicinity of steep shock profiles. Empirically-determined flux limiters are imposed to produce agreement with experimental results in such cases. These flux limiters are numerical factors that reduce the asymptotic free-streaming heat flux, the hypothetical flux that would occur if energy was transported uniformly by all thermal electrons travelling at a characteristic thermal velocity $v_{th} = \sqrt{k_B T_e / m_e}$. The relationship between the free-streaming heat flux and the flux calculated by Spitzer-H{\"a}rm models is such that there is a positive relationship between flux limiter value $f$ and the total flux. This treatment allows Spitzer-H{\"a}rm models to restrain excessive heat fluxes without becoming wildly unphysical because only a small fraction of the total heat flux is carried by electrons approaching the free-streaming velocity in reality\cite{drakeHighEnergyDensityPhysics2018}. Results from a range of flux limiter values are presented in Sec. IV C.

If the temperature gradient scale length becomes too small, even flux-limited Spitzer models break down. As an alternative, xRAGE features a non-local heat conduction model for electrons which employs a Schurtz algorithm\cite{schurtzNonlocalElectronConduction2000}. This procedure goes beyond the Spitzer-H{\"a}rm approximation by discretizing electron energies into a finite number of groups and tracking them separately, enabling a more physical solution near strong electron temperature gradients. The result is similar to that of a flux limiter, in that the peak current is dampened at sharp gradients. Unfortunately, implementing this model was found to be computationally prohibitive for our 2D simulations, but we intend to utilize it in future simulations with improved optimization. Nonetheless, the low variation of our results over a wide range of flux limiters indicates that we are within the regime of validity for flux-limited Spitzer models.

\section{Benchmarking and Transport Options}
\begin{figure*}
\includegraphics{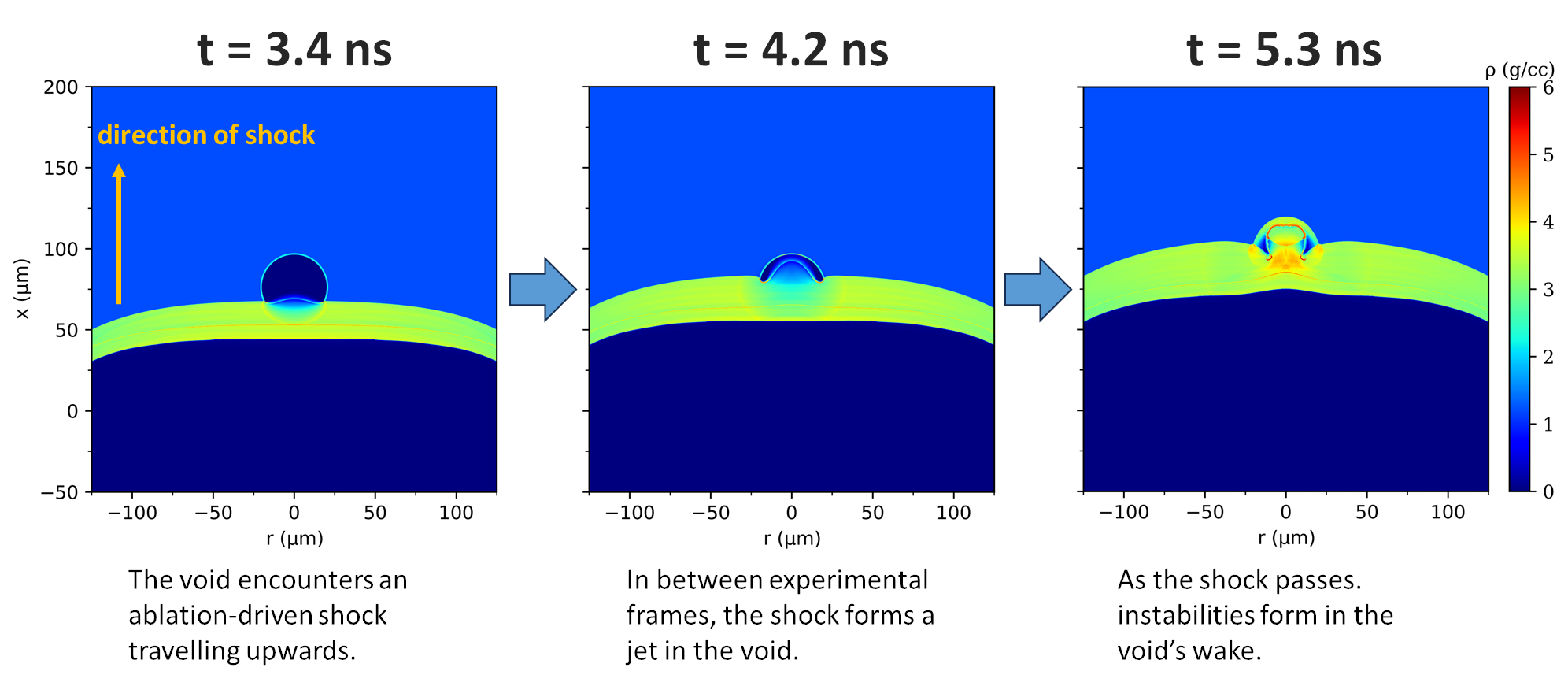}
\caption{\label{fig:fig_2}2D xRAGE simulation of the SBI at various stages of collapse. The shock first reaches the void in the first frame (left). In the center frame, the part of the shock travelling through the void accelerates ahead of the unperturbed shock due to the lower acoustic impedance of the void relative to the surrounding SU-8, forming a plasma jet. In the final frame (right), the jet impulsively displaces the glass shell material on the far side of the void, partially reflects back into the void, and transmits a spherical shock upwards into the SU-8. Instabilities also begin to develop behind the former void. Experimentally, only images of the stages depicted in first and third frames were collected; the simulated intermediate stage aids interpretation of the final stage by showing how it evolves from the first stage.} 
\end{figure*}

\begin{figure*}
\includegraphics{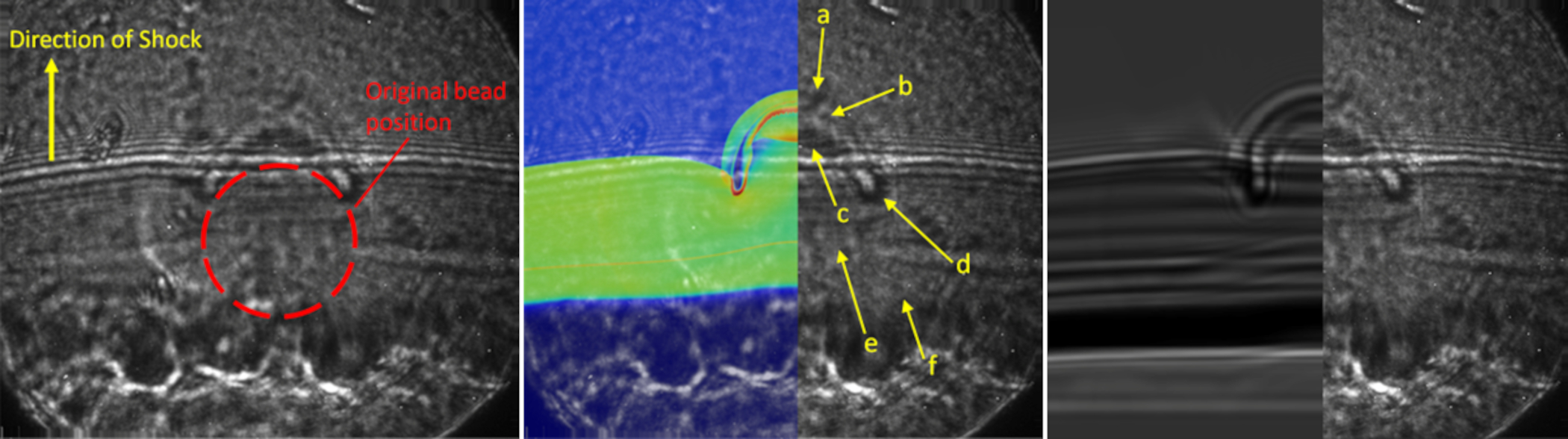}
\caption{\label{fig:fig_3}Left: Experimental XPCI image (single-frame, X-ray energy 17 keV) of a 300 GPa shock collapsing a 42 $\mu$m diameter shell, used as a synthetic void. Center: Comparison of the experimental image with a simulated density map. The simulation reproduces various features in the experimental image that would be otherwise difficult to identify: a) Spherical transmitted shock through the void; b) Impulsively displaced downstream edge of the glass shell; c) Reflected shock from plasma jet impacting the downstream shell edge; d) Lobes of accumulated shell material where the shell surface transitions from convex to concave relative to the oncoming shock; e) Coherent forward-moving Aluminum layer; f) Ablation front. Right: Comparison of synthetic XPCI image forward-modeled from simulated density map with experimental image.}
\end{figure*}

We employ several methods to benchmark our simulations with the experimental results reported by Hodge et al.\cite{hodgeMultiframeUltrafastXray2022} First, our simulated density maps (Fig.~\ref{fig:fig_2}) are used to forward-model synthetic XPCI images, and both are then compared with XPCI images of the experimental void at equivalent stages of collapse to establish qualitative agreement (Fig.~\ref{fig:fig_3}). This comparison shows that our simulations recreate the main features observed in the X-ray images. Because the development of simulated features can be tracked at much higher framerates than is possible experimentally, this agreement allows us to characterize features in the X-ray images that would otherwise be difficult to interpret. In particular, our simulations show the development of a plasma jet due to the acceleration of the initial shock as it enters the low-density void. When this jet impacts the far side of the void, a spherical shock is transmitted into the SU-8, and a reflected shock is produced. These shocks are observed both experimentally and in our simulations. The downstream shell material is also impulsively displaced by the plasma jet, which explains why the shell appears to move ahead of the unperturbed shock in the experimental images. Quantitative comparison of synthetic and real images is beyond the scope of this work, but will be the subject of an upcoming paper.

For quantitative benchmarking, we compare our simulated shock velocities to experimental shock timings derived from multi-frame UXI images. The acquisition of multiple images from a single shock-compressed sample allowed us to directly measure the shock velocity while minimizing the uncertainty due to shock-to-shock variations. Our procedure for identifying a shock position and pressure in the simulated pressure fields, and the challenges involved in doing so, are discussed in Sec. IV A. It is important here to note that our simulations represent a 2D planar slice through the center of the target, while the UXI images reflect the total absorption of the X-ray imaging beam along its path through the target; furthermore, the 42 micron diameter shell makes up only 10\% of this path. Therefore, it is appropriate to perform our benchmarking analysis on the unperturbed shock outside the region of the void. We also compare simulated and experimental shock pressures. This procedure is carried out on simulations with radiation turned off, gray radiation diffusion enabled, multigroup radiation diffusion enabled, and gray IMC radiation enabled (Sec. IV B), and with various heat conduction flux limiters (Sec. IV C). 

\subsection{\label{sec:level4}Defining Simulated Shock Pressure and Location}
Associating a location and pressure with a simulated shock is non-trivial, in part because the cell size of the computational grid is smaller than the width of the shock front. In simulations of experiments with a single well-defined shock, it may be sufficient to identify the cell with the maximum pressure derivative $\frac{\partial P}{\partial x}$ in the direction of shock propagation as the location of the shock. More complicated pressure profiles including multiple shocks may necessitate the use of the logarithmic pressure derivative $\frac{\partial\left(\log{P}\right)}{\partial x}$, i.e., the largest order-of-magnitude change in pressure over a cell, to identify the primary shock. This method requires exceptions for initial conditions in which discontinuous material boundaries create pressure jumps over a single cell that cover a larger range of orders of magnitude than the shock itself. If identification of secondary shocks is desired, or if precursor shocks are present ahead of the primary, a piecewise scheme in time and space using one of these methods can be constructed based on the expected order and location of discontinuities of interest. In the present work, which examines a single primary shock that consistently leads all other pressure waves, a global maximum derivative method is used. Absolute and logarithmic derivatives were found to give identical results over almost all times, with fewer than $1\%$ of times producing a one-cell (0.1 micron) offset between the two derivatives' maxima. The simpler absolute derivative method is thus used in this work.

Selecting a cell in which to measure the shock pressure presents an additional problem. The physical half-width of the shock front is several hundred nm, while the computational resolution is 100 nm. The result of this is that the pressure associated with the shock occurs several cells or more upstream (i.e., in the opposite direction of the shock's propagation) of the cell containing the steepest pressure gradient, and there is a large pressure difference between these cells precisely due to this large gradient.

\begin{figure}
\includegraphics[width=0.49\textwidth]{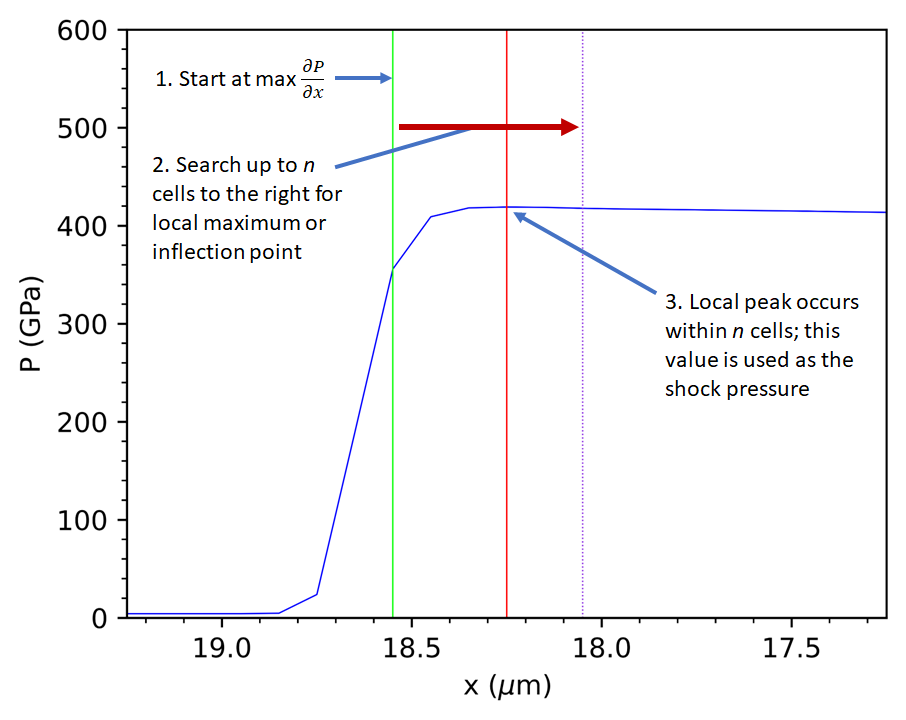}
\caption{\label{fig:fig_4}Depiction of algorithm used to select shock pressure with an example pressure profile. Beginning at the location of the maximum pressure gradient, denoted by the green (leftmost) vertical line, the algorithm moves to the right searching for either a local pressure maximum or an inflection point in the pressure profile, which is denoted by the central red vertical line. If such a point has not been found after searching $n$ cells, the search terminates and uses the pressure at the current point (denoted by the vertical dotted line). This gives consistent results if $n$ is chosen such that the search region is within about a micron of the physical shock front width.}
\end{figure}

To resolve this, we implement the following algorithm, which is depicted visually in Fig.~\ref{fig:fig_4}. For an idealized shock moving in the negative x direction, the pressure gradient decreases monotonically in between the location of the maximum pressure gradient and the location of the peak pressure, where the gradient reaches zero. This local pressure maximum is the pressure associated with the shock. However, at some times there is a secondary pressure wave or shock in the process of merging with the primary, which may cause the pressure gradient to begin increasing again before it falls to zero. In this case, the inflection point where the derivative of the pressure gradient goes from negative to positive can be used as the "top" of the primary shock. Accordingly, we select the shock pressure by working forward (in the positive x direction, opposite the shock's propagation) from the location of the maximum pressure gradient, and upon encountering either a local pressure maximum or an inflection point, evaluate the pressure in that cell. Rarely, there will be a pressure waveform on top of the shock front such that neither the first nor second pressure derivative reaches zero; the pressure gradient continuously decreases from its maximum, but at a reduced rate. To handle these pathological cases, we implement a hard cut-off value $n$ such that, upon searching $n$ cells behind the maximum pressure gradient without finding a maximum or inflection point, the pressure in this $n$th cell will be taken as the shock pressure. This cut-off is necessary because the real shock spans only about half a micron, and should be chosen such that $n$ cells represent this distance. In our simulations, we find physically reasonable values of $n$ (within about 1 micron of the actual shock width) give identical results.

\subsection{\label{sec:level5}Radiation Transport}
Initial modeling for the experiment carried out by Hodge et al.\cite{hodgeMultiframeUltrafastXray2022} did not consider radiation transport. Nonetheless, these simulations showed good qualitative agreement with experimental results and are reproduced here in Fig.~\ref{fig:fig_2}. Generally, the presence of radiation can be expected to result in weaker shocks. This is because the entire laser-target system is essentially a rocket motor: the laser irradiates the Kapton layer, which is explosively ablated, and the target experiences an impulsive force opposite the velocity of the ablated plasma. The strength of the shock increases with the velocity of the ablated plasma due to momentum conservation. In the absence of radiation transport, the energy of the laser drive is converted into the kinetic and thermal energy of the ablated plasma and the shocked target; when radiation transport is present, it provides an additional pathway for this energy, resulting in lower ablation velocity and pressure.

\begin{figure}
\includegraphics[width=0.5\textwidth]{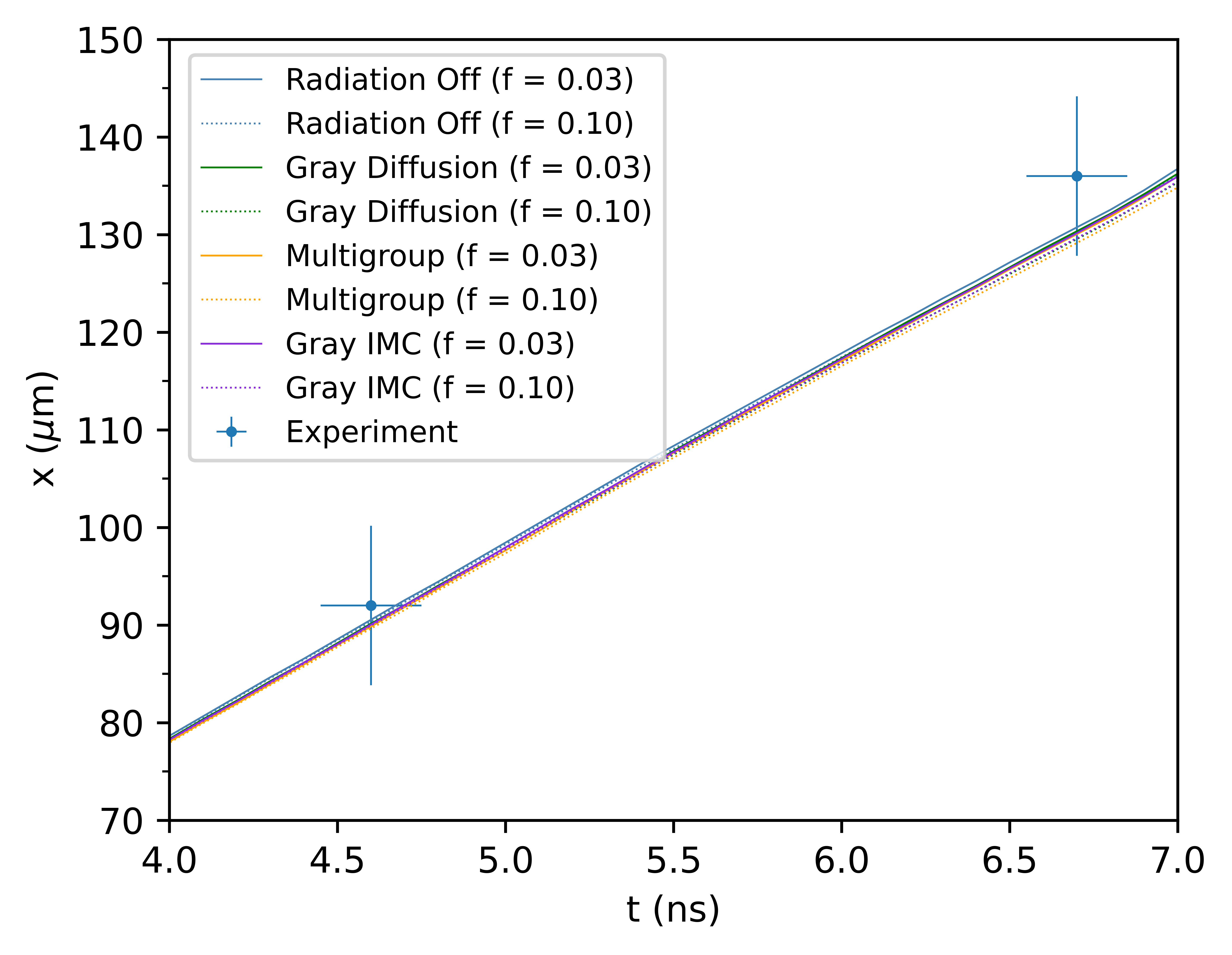}
\caption{\label{fig:fig_5} Shock position as a function of time with different radiation options and flux limiters. Flux limiters of 0.03 and 0.10 are used (see Sec. IV C). Shock positions without radiation transport, with single- and multi-group gray radiation diffusion, and with gray IMC radiation are shown. The spread in simulated shock positions is negligible. In all cases, the average shock velocity is 13\% lower than what is observed experimentally, but is within measurement error of the experimental data (Figs. ~\ref{fig:fig_6} and ~\ref{fig:fig_7}).}
\end{figure}

In the present study, we investigate the role of radiation transport in faithfully reproducing experimentally observed shock pressures. The empirical pressure is calculated using the shock velocity obtained from known shock positions at different times and a polyamide equation of state\cite{moreNewQuotidianEquation1988,youngNewGlobalEquation1995}. Therefore, the most direct comparison between experiment and simulation for benchmarking purposes is of these shock positions. Results are presented from simulations conducted without radiation transport enabled, with flux-limited single- and multi-group gray radiation diffusion, and with a gray IMC radiation model. We consider the shock properties one shell diameter, or 42 microns, from the axis of symmetry. This distance is chosen so that we are sampling the shock where it is not perturbed by the shell or void during the experimental window (as discussed in the beginning of this section), while still remaining reasonably planar. The position of the shock as a function of time is plotted for each radiation option and compared with experimental values in Fig.~\ref{fig:fig_5}. The experimental positions are determined from the location of the Fresnel fringes\cite{montgomeryInvitedArticleXray2023} associated with the shock in the UXI images, which are known to deviate from the "true" shock location as defined by the pressure gradient. As such, a small, empirically-determined correction (0.13 µm) is applied to the experimental shock positions to compensate for this drift. 

While there is almost no spread in shock position with different radiation options, there is a consistent relationship over the experimental window: at any given flux limiter value, the shock has traveled furthest in the case that radiation transport is not modeled, as expected, and the shocks modeled using single-frequency radiation schemes are ahead of those modeled with multi-group radiation. The gray IMC model is almost congruent with the gray diffusion model, but neither is consistently ahead of the other. This relationship also holds at earlier times.

In all cases, the simulated shock lags behind the real shock, but is within experimental error. This error arises from several sources. First, there is a $\pm 0.83$ µm uncertainty in the measured shock position in the UXI images due to pixel size, as well as a time uncertainty of $\pm 0.15$ ns. This time uncertainty is systematic and is due to ambiguity in the arrival time of the X-ray pulse train used to capture the UXI images. There is also an uncertainty in the simulated shock positions due to the finite computational grid size and time step, which are 0.1 µm and 0.1 ns respectively. The largest spatial error comes from uncertainty in aligning the coordinate grid of the experimental images with the computational grid. This is done by comparing the location of static material boundaries at $t < 0$, before the laser turns on. However, while these boundaries are essentially one-dimensional in reality and are modeled as such in our simulations, they are smeared out over 6.5 microns in the UXI images due to physical imaging limitations. Finally, there is an additional 1-pixel uncertainty in the measured shock position due to potential misalignment of the shock propagation vector and the ablator surface, which we estimate to be $\pm 1$ degree. The total error in absolute shock position, relative to the simulated values, is $\pm 6.6$ µm and $\pm 0.18$ ns. However, this is overwhelmingly due to systematic errors affecting both position measurements equally, and therefore irrelevant to measurements of shock velocity.

\begin{figure}
\includegraphics[width=0.5\textwidth]{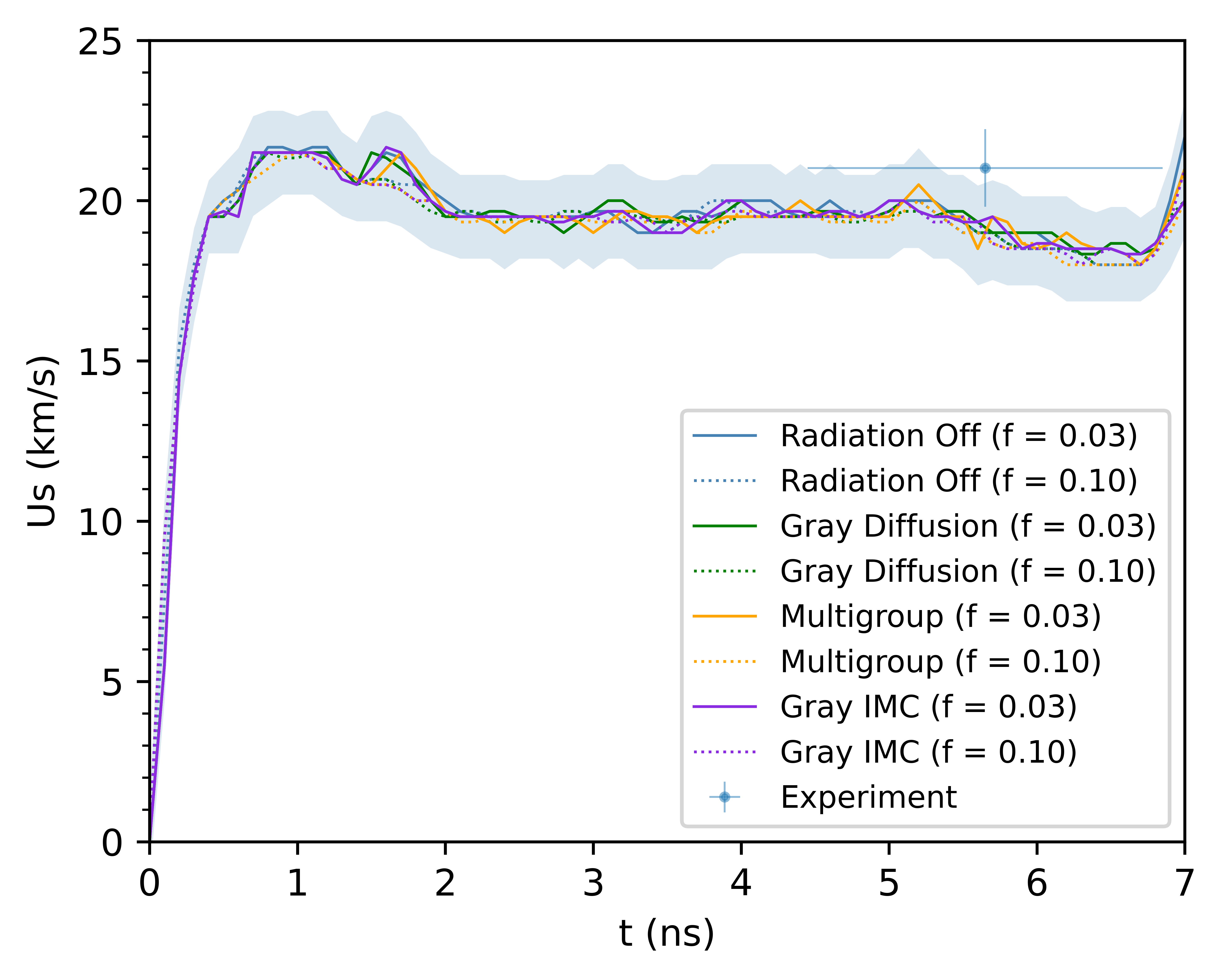}
\caption{\label{fig:fig_6} Shock velocity as a function of time with different radiation options and flux limiters, compared with the experimentally-measured shock velocity. The shaded region around the simulated velocities represents the error due to computational grid resolution and time-step size. Because the experimental value is an average, it is portrayed as an instantaneous velocity with horizontal error bars encompassing the averaging window. The time uncertainty on the experimental shock positions from which this velocity is derived is much smaller ($\pm0.15$ ns, as in the previous figure).}
\end{figure}

\begin{figure}
\includegraphics[width=0.5\textwidth]{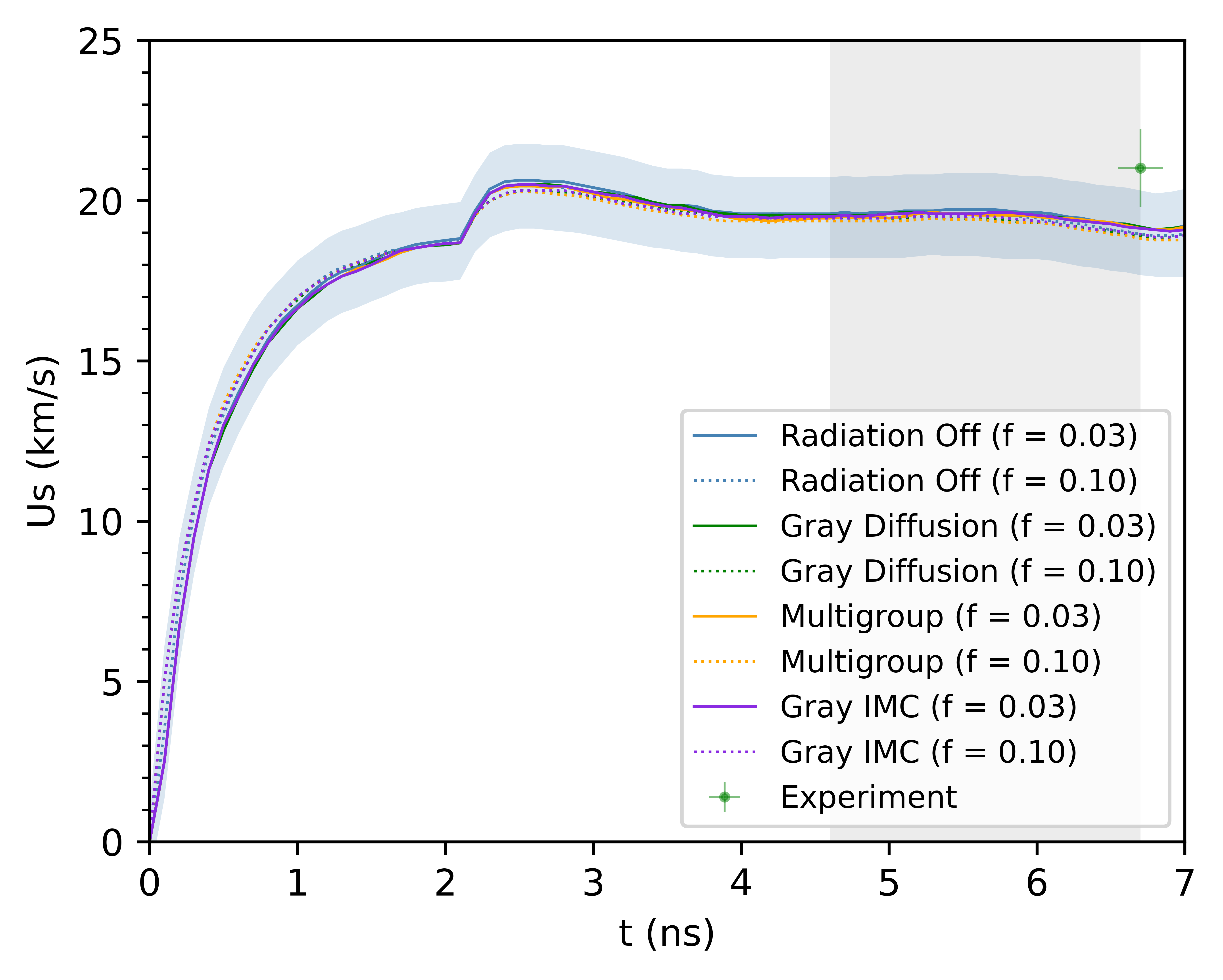}
\caption{\label{fig:fig_7} Rolling time-averaged shock velocity as a function of time. The value at each time is the average instantaneous velocity over the previous 2.1 ns, the size of the experimental window. This window is represented by the shaded rectangular region between 4.6 and 6.7 ns. The experimental timing error is almost entirely due to uncertainty in the arrival time of the X-ray pulse train, which affects all time measurements equally. Error in the time between the shock measurements is due to uncertainty on the time between individual pulses, which is negligible ($\pm$25 fs).}
\end{figure}

The slopes of the lines in Fig.~\ref{fig:fig_5} are the instantaneous shock velocities, and are compared with the velocity derived from the shock displacement over the experimental window in Fig.~\ref{fig:fig_6}. Fig.~\ref{fig:fig_7} shows a rolling time average of the simulated shock velocities over a 2.1 ns window, such that the value at 6.7 ns is directly comparable to the experimental velocity. This comparison shows that our simulations reproduce experimental shock velocities to within measurement error in all cases. Here the 6.5 micron grid alignment uncertainty in the absolute shock positions is irrelevant, because we are concerned only with the difference between the two positions; the remaining uncertainties in shock displacement and velocity are given in Table~\ref{tab:table2}. The instantaneous shock velocities highlight the differences between each radiation option at any given point in time, but obscure the correlations over time revealed by the absolute positions. There is no discernible general relationship between instantaneous shock velocity and radiation scheme--the correlations only become apparent when the velocity is averaged or integrated over time. However, even here the variation with radiation scheme is negligible.

Table~\ref{tab:table2} shows the average shock velocity over the experimental window for each radiation option, along with shock pressures calculated directly from these velocities using a polyamide equation of state\cite{moreNewQuotidianEquation1988,youngNewGlobalEquation1995}. This comparison is the most accurate for benchmarking purposes, because it removes variations due to grid alignment and systematic measurement errors, and further reinforces that our simulations accurately reproduce the empirically observed shock velocity. However, it provides much more limited information on the general effect of radiation model (and heat conduction) on the shock. The calculated pressure is provided to contextualize the velocity measurements, but should not be used to evaluate simulation fidelity due to its dependence on equation of state (both in outright value and uncertainty) and, for the experimental measurement, sensitivity to small errors in the density of the unshocked SU-8. It is calculated as a function of shock speed from a given equation of state and initial density; while in reality (and in simulations) small changes in initial density cause small changes in shock pressure and velocity, small deviations in the empirical value of the un-shocked SU-8 significantly change the calculated shock pressure, because the empirically-measured shock velocity is fixed.

It should also be noted that equation of state choice affects the simulated shock speed to some degree. The rolling time-averaged shock speed predicted by the CH equation of state from SESAME is ~0.1 km/s higher than that of the polyamide equation of state from LEOS over the experimental window. However, this variation is extremely small relative to experimental uncertainties, and only causes a 1-2 um shift in shock position by t = 7 ns. Our simulations were similarly insensitive to whether or not the tabulated equation of state values were scaled to average atomic number $\bar{z}$ and mass number $\bar{a}$, which suggests that deviations due to uncertainty in tabulated equation of state values are not significant.

\begin{table}
\caption{\label{tab:table2}Shock displacement over the 2.1 ns experimental window and derived shock pressures using a LEOS polyamide equation of state\cite{moreNewQuotidianEquation1988,youngNewGlobalEquation1995}. Results from various radiation options with heat conduction flux limiters of 0.03 and 0.10 are shown. Errors on experimental values are due finite pixel size ($0.83$ µm), while errors on simulated values are due to computational grid resolution and time step size.}
\begin{ruledtabular}
\begin{tabular}{cccc}
\mbox{Simulation} & $\Delta x$ & \multicolumn{1}{c}{$\Delta x/\Delta t$} & \mbox{P} \\
 & $\mu\mathrm{m}$ & \multicolumn{1}{c}{$\mathrm{km/s}$} & $\mathrm{GPa}$ \\
\hline
\mbox{Experiment} & $44.01\pm2.3$ & $20.96\pm1.1$ & $341\pm45$\\
No Rad & $40.2\pm0.14$ & $19.14\pm1.3$ & $298\pm46$ \\
No Rad & $39.8\pm0.14$ & $18.95\pm1.3$ & $292\pm46$ \\
Gray Dif & $40.2\pm0.14$ & $19.14\pm1.3$ & $298\pm46$ \\
Gray Dif & $39.6\pm0.14$ & $18.86\pm1.3$ & $289\pm46$ \\
Mgd & $40.2\pm0.14$ & $19.14\pm1.3$ & $298\pm46$ \\
Mgd & $39.5\pm0.14$ & $18.81\pm1.3$ & $287\pm46$ \\
Gray IMC & $40.1\pm0.14$ & $19.10\pm1.3$ & $297\pm46$ \\
Gray IMC & $39.7\pm0.14$ & $18.90\pm1.3$ & $290\pm46$ \\
\end{tabular}
\end{ruledtabular}
\end{table}

\begin{figure}
\includegraphics[width=0.5\textwidth]{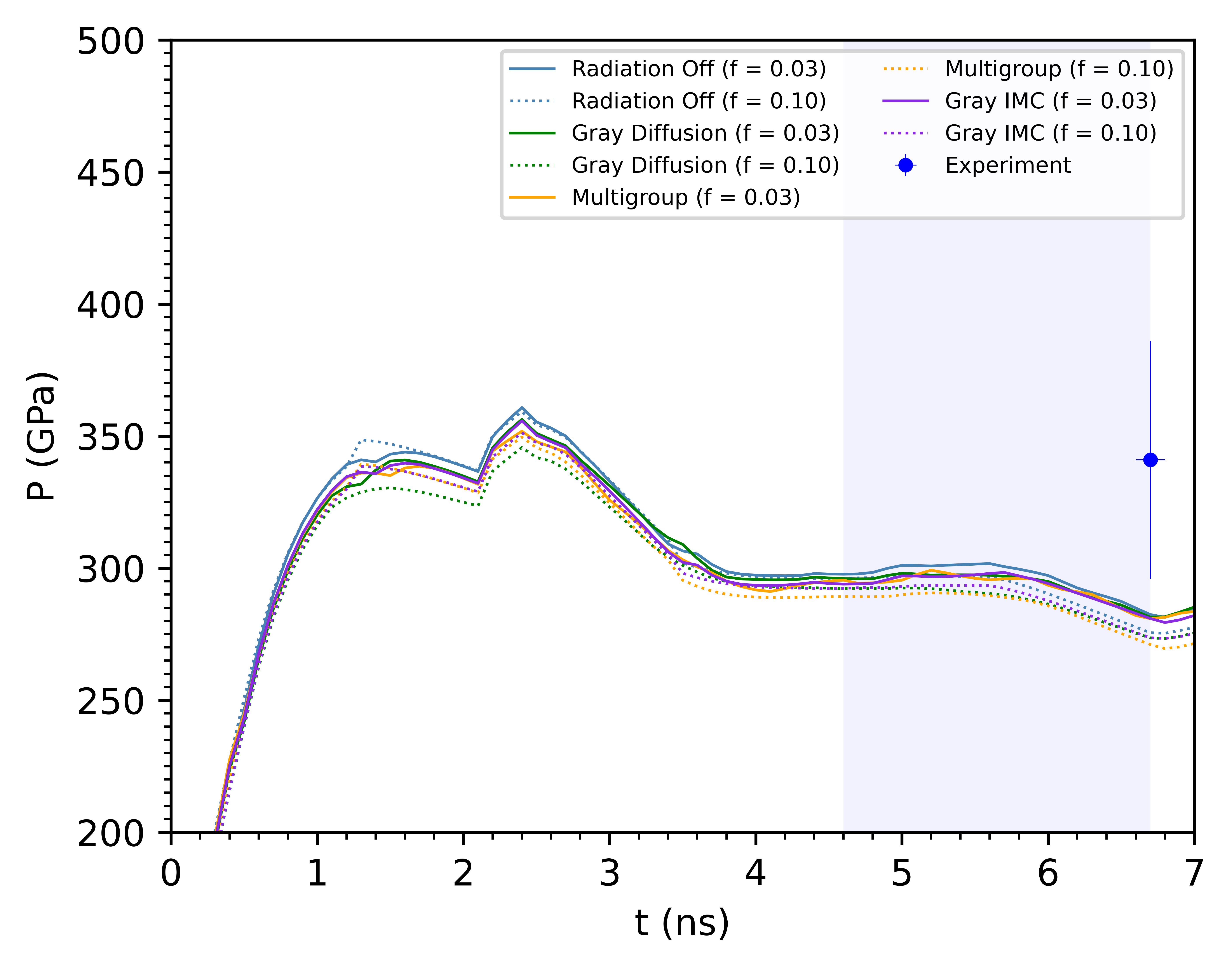}
\caption{\label{fig:fig_8} Simulated rolling time-averaged shock pressures with various radiation options and heat conduction flux limiters. The experimental pressure value was obtained from the average shock velocity from 4.6 ns to 6.7 ns, which is indicated by the shaded region in the plot. The averaging window for the simulations is 2.1 ns, corresponding to the time between experimental frames.}
\end{figure}

Despite its shortcomings as a benchmarking metric, shock pressure is the more useful quantity to consider when evaluating the effect of radiation model on the system in general. This is because it varies considerably with even small changes in shock speed. The experimental pressure value from Table~\ref{tab:table2}, which represents a time-average over the experimental imaging window, is compared with the rolling time-averaged shock pressure in our simulations in Fig.~\ref{fig:fig_8}. The simulated pressure is averaged over the previous 2.1 ns, as in Fig.~\ref{fig:fig_7} so that it represents an average over the experimental window when evaluated at the experimentally-sampled time. We find that the average shock pressure decreases (as expected) by up to 20 GPa when radiation is enabled. A general relationship between single- and multi-group radiation models cannot be given, because the multiple opacities of the multi-group model couple non-trivially to different heat conduction flux limiters (Sec. IV C). The more physical IMC model produces pressures up to 10 GPa higher than the diffusive models, but this difference is only a few percent of the total shock pressure.

\begin{figure}
\includegraphics[width=0.5\textwidth]{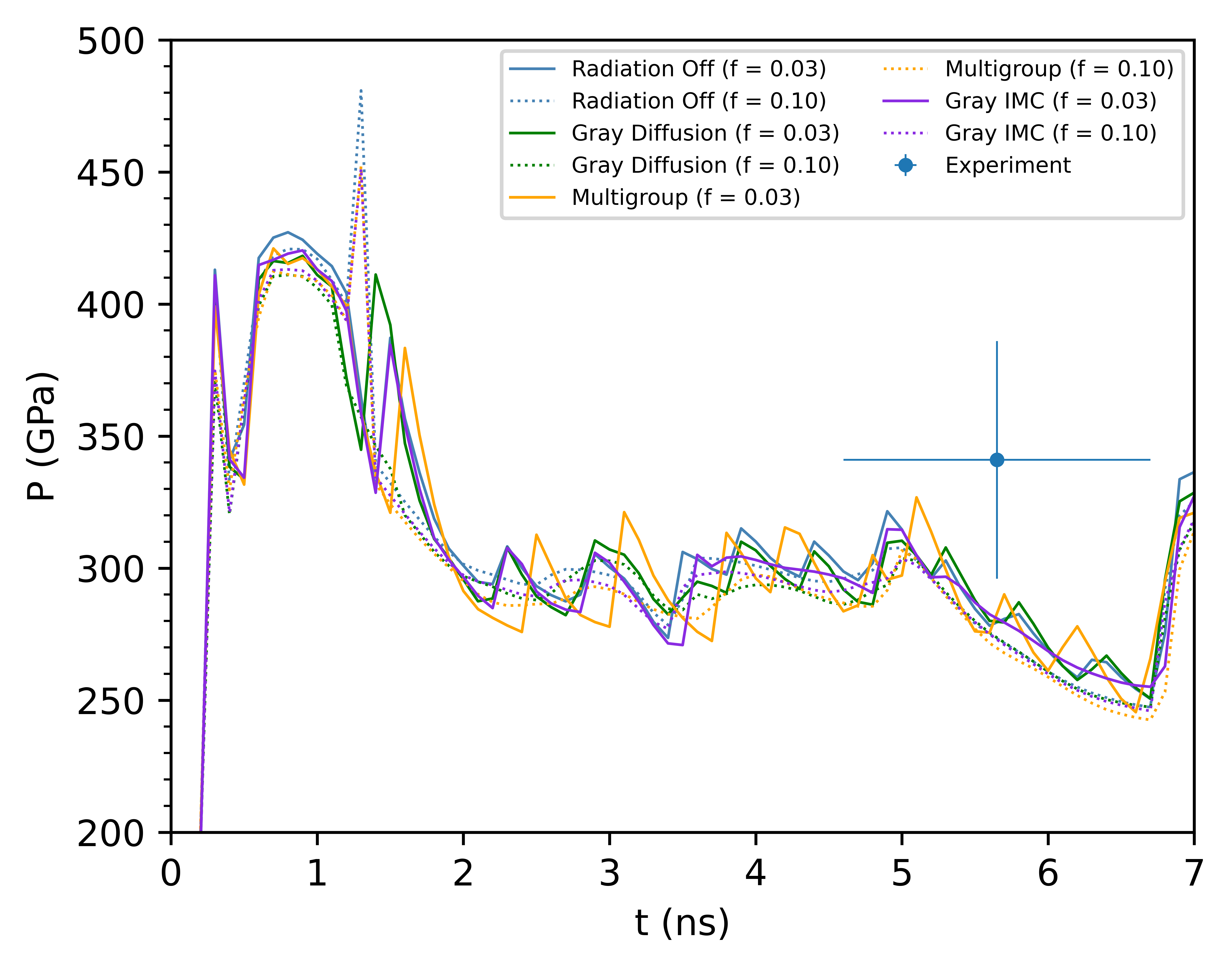}
\caption{\label{fig:fig_9} Simulated instantaneous shock pressure as a function of time. Results from different radiation models and heat conduction flux limiters are shown. As in Fig.~\ref{fig:fig_6}, the horizontal error bars on the experimental value represent the uncertainty in assigning a precise time to an average value, rather than uncertainties in the measurements used to produce this value.}
\end{figure}

When considering the instantaneous shock pressure (Fig.~\ref{fig:fig_9}), the situation becomes more complicated. The relationship between radiation option (and even whether radiation is enabled at all) and shock pressure varies unpredictably with time. This unpredictability decreases with higher flux limiters, but even at the maximum value examined ($f = 0.1$), there is not a consistent hierarchy across the time domain. Nonetheless, Fig.~\ref{fig:fig_8} demonstrates that the average shock pressure is always lower when radiation is modeled.

\begin{figure}
\includegraphics[width=0.5\textwidth]{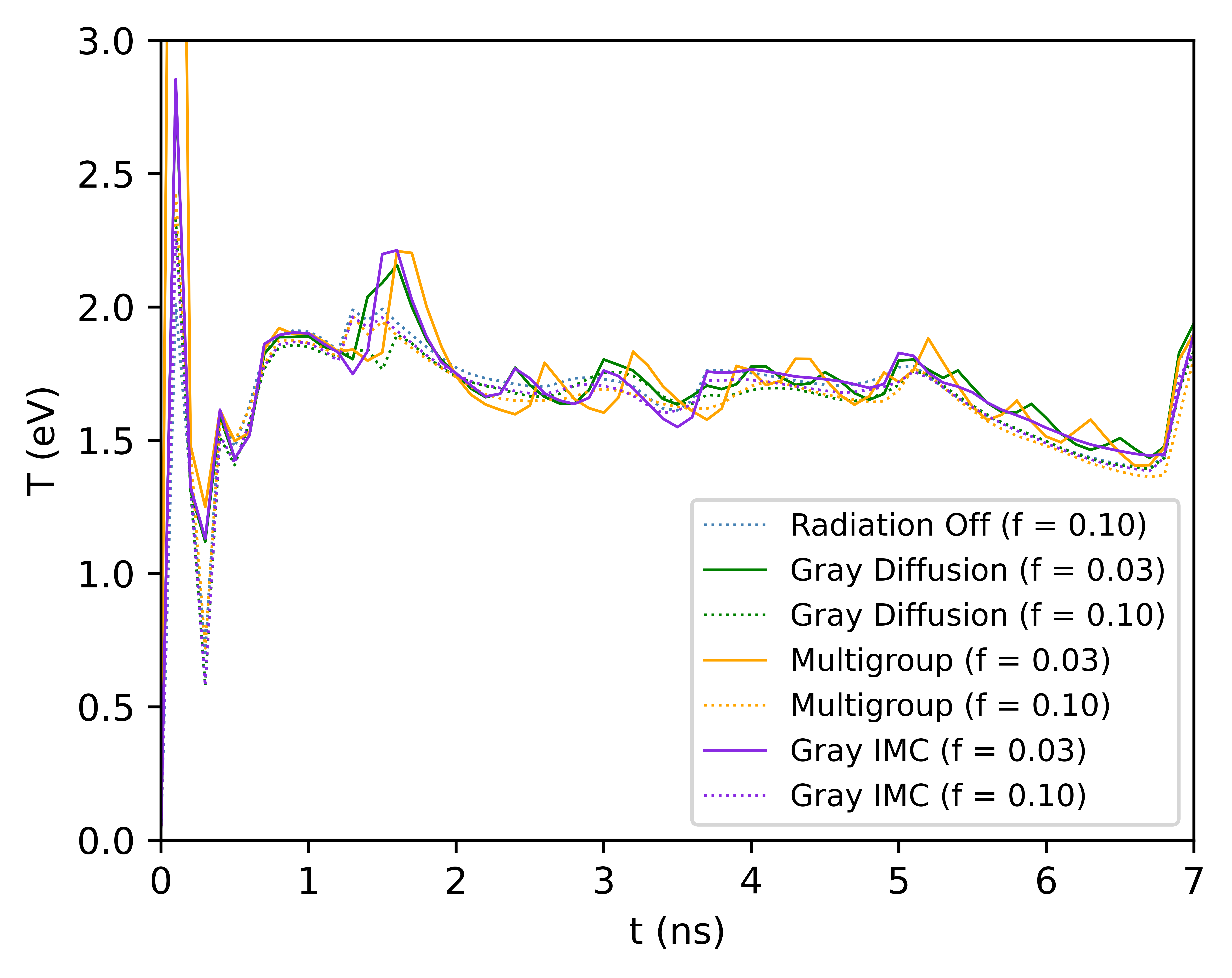}
\caption{\label{fig:fig_11} Shock temperatures with different radiation models and flux limiter values.}
\end{figure}

We also note that enabling radiation transport models results in radiative preheating of the glass shell. Our simulations indicate that the temperature of the ablated plasma is $\SI{6e6}{\kelvin}$ (Fig.~\ref{fig:fig_10}); the distance between this plasma and the glass shell is never more than 2.5 mean free paths for thermal X-rays at this temperature, so some preheating of the shell is expected. However, our simulations indicate that the shell remains well below (< 20\%) the melting temperature of glass, and therefore that this preheating does not have a discernible effect on the system after the shock arrives. The shock itself is weakly radiative, but has a much lower temperature of $\SI{4.6e5}{\kelvin}$ (Fig.~\ref{fig:fig_11}). The mean free path of thermal radiation from the shock is no more than a few microns, so there is virtually no preheating of the shell or SU-8 from this source.

\subsection{\label{sec:level6}Heat Transport}
As discussed in Sec. III B, this work uses a local Spitzer conduction model for electron heat transport. Such models use flux limiters to restrain otherwise-unphysical heat flows that occur when the mean free path of electrons approaches the scale length $T_e/\nabla T_e$. ICF studies have shown that appropriate flux limiter values for HED shock simulations range from 0.025 to 0.150\cite{backMeasurementsElectronTemperature1996}, with hohlraum simulations typically using a value of 0.050.

We examined flux limiter values $f$ from 0.03 to 0.10. This value is a multiplier applied to the asymptotic free-streaming heat flux at each time step, which in Spitzer models is positively related to the total heat flux, so a higher flux limiter will allow for enhanced diffusion of heat from regions of high temperature. Assuming that the simulated system is in the regime of Spitzer validity, the choice of $f$ can be expected to affect the measured shock pressure in several ways. Two of these are due to the physics of laser ablation described at the beginning of this section. After the initial ablation first generates a coronal plasma, subsequent laser energy is deposited where this plasma reaches the critical density. This ablation launches the primary shock into the target and then begins to accelerate the target via the rocket effect, as discussed above in Sec. IV B. After the first few hundred picoseconds, the ablation front moves forward past the critical surface, and laser energy must be conducted through over-dense plasma to reach the ablation front. This conduction occurs overwhelmingly through heat transport by thermal electrons. The resulting pressure at the ablation front, which drives the primary shock, is directly proportional electron temperature $T_e$ at the critical density. Since the critical density is a function only of frequency, at a given laser intensity and frequency $T_e$ is inversely related to $f$\cite{drakeHighEnergyDensityPhysics2018}:
\begin{equation}
P_{abl}\propto k_B T_e = \left(\frac{a}{f}\frac{I_L \sqrt{m_e}}{n_c}\right)^{2/3}.
\end{equation}
Here $I_L$ is the laser intensity, $n_c = \frac{m_e}{4\pi e^2}\omega^2$ is the critical density, and $m_e$ is the electron mass. $a$ is the fraction of the deposited laser energy that is conducted to the target; virtually all of the remaining laser energy ($\left(1-a\right)I_L$) drives the expansion of the coronal plasma. If $a$ is assumed to be constant, at higher values of $f$ more heat is conducted away from the coronal plasma. This lowers $T_e$ and therefore the ablation pressure driving the shock.

However, the value of $a$ is likely dependent on $f$. If $f$ is increased, more of the total laser energy can be conducted away from the coronal plasma, which translates to a higher value for $a$. Therefore, both the numerator and the denominator in Eqn. (12) grow with $f$, and the overall effect depends on which effect is stronger. It is also important to note that while the ablation pressure drives the shock, it is not generally equivalent to the shock pressure. Changes in pressure at the ablation front take time to propagate to the shock front, and energy from the laser has to be further conducted from the ablation front through the shocked target to reach the shock. To first order, the effectiveness of this process also has a positive dependence on $f$. However, it is also affected by the presence of secondary shocks in the region between the ablation front and the primary shock. Shock fronts themselves dissipate energy more rapidly with higher $f$, and smaller secondary shocks may not form at all with sufficient heat conduction. If the effect of reflected shocks and rarefaction waves in the shocked region on the primary shock is small, then at early times, when the shock is impulsively-driven, the explicit dependence of $T_e$ on $f^{-2/3}$ in Eqn. (12) is likely most influential on shock pressure. At later times, when heat has to be conducted through more of the shocked target to reach the shock front, a positive relationship between shock pressure and $f$ may develop.

In practice, our simulations indicate that secondary shocks in the accelerated target have a large effect on the experimentally-measured shock pressure, and are in fact the dominant channel by which the value of $f$ affects the primary shock. Forward-moving shocks in this region arise periodically in the following manner: a forward-moving shock (initially the primary shock) is partially reflected as it moves from the Kapton layer to the higher-impedance aluminum. This reflected shock travels backwards (upstream) until it reaches the ablation front. The ablated material upstream of the ablation front has a lower impedance than the shocked Kapton, so a reflected rarefaction wave is produced, which travels forward some distance before dissipating. The ablation pressure then re-compresses the rarefied Kapton, producing the forward-moving secondary shock. As these shocks encounter the aluminum layer, they are partially reflected, and the process repeats.

When these secondary shocks reach the initial shock, they momentarily increase the shock velocity and pressure, causing the regular spikes from 2 ns onward in Fig. ~\ref{fig:fig_6} and, more noticeably, Fig. ~\ref{fig:fig_9}. These spikes elevate the time-averaged velocity and pressure (which are what is calculated experimentally) when they occur within the averaging window. In our simulations, higher values of $f$ effectively suppress these shocks, resulting in lower average velocities and pressures as shown in Figs. ~\ref{fig:fig_7} and ~\ref{fig:fig_8}. Over the times sampled experimentally, these spikes are almost entirely responsible for the differences in shock strength between different $f$. Conversely, during the initial phase of ablation prior to the development of secondary shocks, higher values of $f$ result in marginally higher shock velocities. This is unexpected and may indicate a strong ($>f^{1}$) initial dependence of $a$ on $f$. Nonetheless, the average shock pressure varies by only 10 GPa at most over the range of $f$ examined, and the difference in shock velocity is negligible after the first few hundred picoseconds.

\begin{figure}
\includegraphics[width=0.5\textwidth]{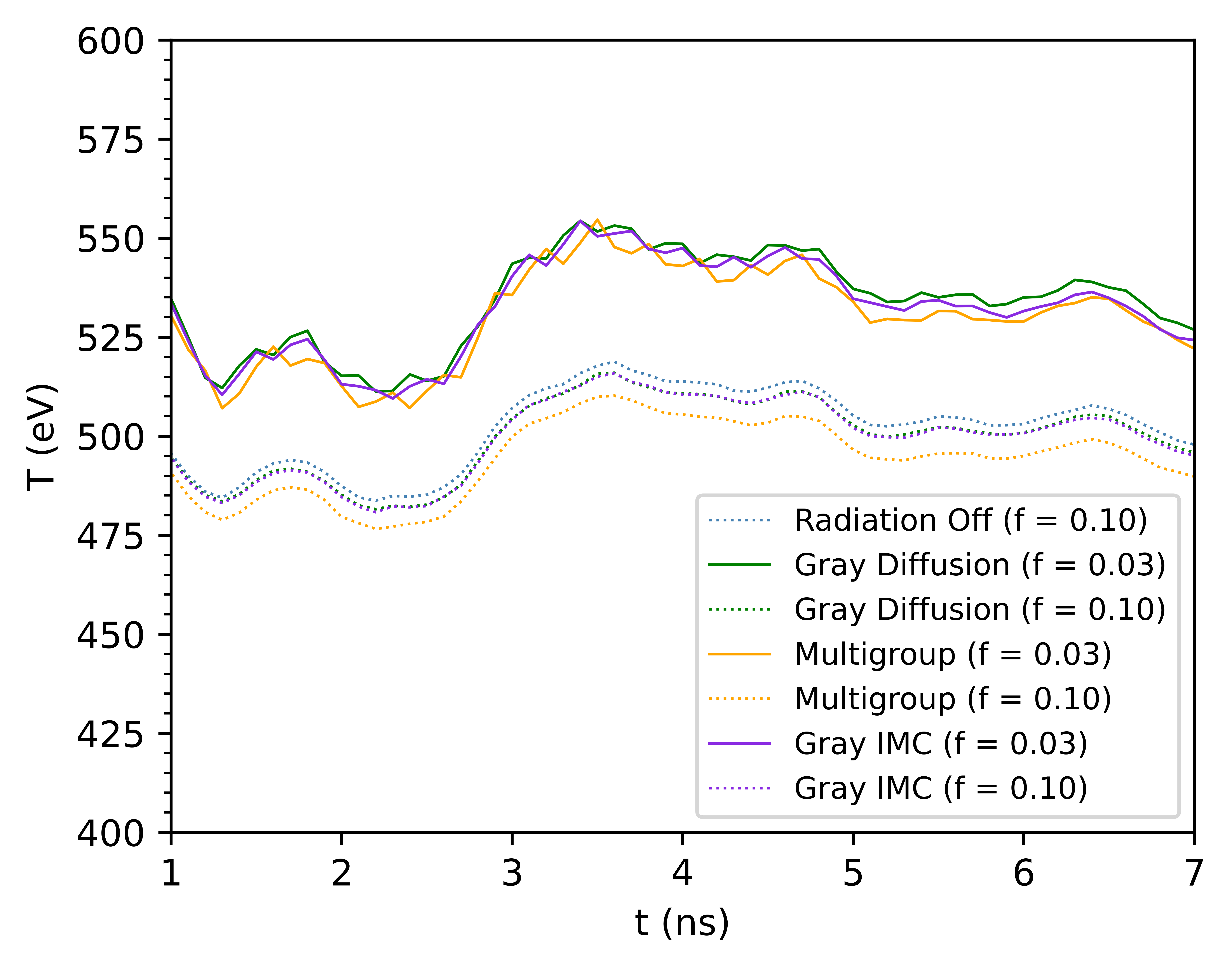}
\caption{\label{fig:fig_10} Temperature of ablated coronal plasma with different radiation models and flux limiter values. Lower (more aggressive) flux limiter values prevent heat from escaping the coronal plasma, resulting in a higher $T_e$ in this region.}
\end{figure}

A final effect of heat conduction is on radiative preheating of the glass shell. As discussed previously, this preheating is due to thermal x-rays emitted by the coronal plasma. Higher values of $f$ allow more heat to flow away from the coronal plasma, lowering its temperature (Fig. ~\ref{fig:fig_10}) and decreasing its thermal emission. Moreover, higher $f$ decrease the range and magnitude of the spurious diffusion associated with this preheating in our simulations. This diffusion arises because our simulations do not include material strength; the target is artificially frozen in place prior to the arrival of the leading shock using the "quiet start" option in xRAGE. While this method is common in HED shock simulations, it does allow some gasdynamic diffusion of heat and pressure prior to shock arrival, which is unphysical -- though, as discussed previously, in our case this does not affect the development of the system after the shock arrives due its negligible magnitude and extent.

\section{Hydrodynamic Instabilities}
\begin{figure}
\includegraphics[width=0.5\textwidth]{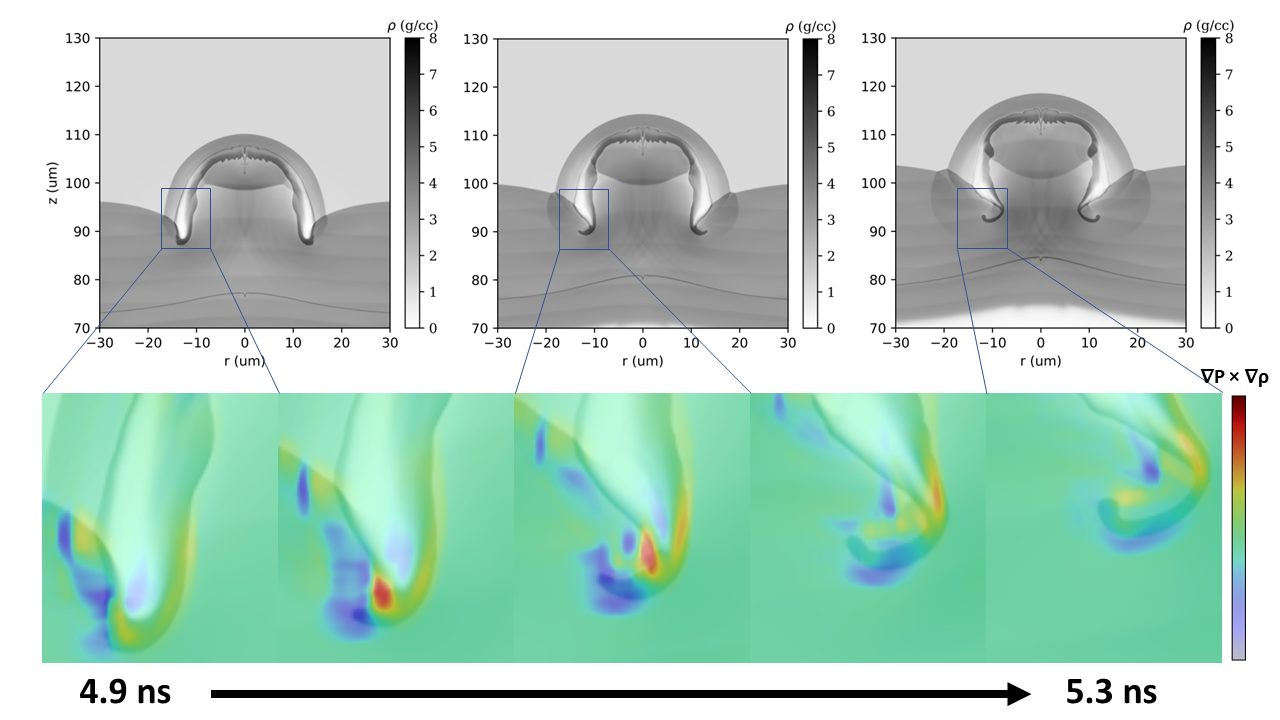}
\caption{\label{fig:fig_12} Simulated baroclinicity. The top row comprises density maps of the developing system, which show the formation of instabilities immediately upstream of the widest part of the shell after the shock passes. The bottom row shows a zoomed-in cut-out of the instability region over time, superimposed on a map of the baroclinicity in this region. The baroclinicity coincides with the instability and is highest when it is growing most rapidly, which supports our expectation that the instability is baroclincally driven.}
\end{figure}

In all our simulations, vorticity generation was observed immediately behind the "lobes" that form between the interior jet and the unperturbed shock (Fig.~\ref{fig:fig_12}, top row).
This vorticity develops into an instability by $t =5$ ns. At conventional pressures, SBIs are known to generate Richtmyer-Meshkov instabilities in which baroclinicity created by an impulsive shock leads to vorticity generation. To confirm whether this was occurring in the present case, we mapped the cross product of the pressure and density gradients in our simulated flow field\cite{zhao2022scale,yin2023energy}, and compared this with our density maps (Fig.~\ref{fig:fig_12}).
This comparison clearly shows that the instabilities observed are driven by baroclinic misalignment between pressure and density gradients, as in the low pressure case, and is in agreement with recent work investigating the role of baroclinicity in laser-driven divergent SBI instabilities\cite{liuDynamicShockBubble2023}.

\section{Conclusions}
We have demonstrated that 2D xRAGE simulations accurately reproduce the qualitative features and quantitative shock characteristics observed in the laser-driven shocked void experiment carried out by Hodge et al.\cite{hodgeMultiframeUltrafastXray2022} These simulations can thus be used to track the development of features present in X-ray images at earlier times and at higher resolutions and framerates than are achievable experimentally. This capability is vital to interpreting images produced in such experiments and will facilitate enhanced quantitative analysis of laser-driven SBI at HED pressures, a regime that has only recently begun to be probed experimentally and is of crucial importance to inertial fusion energy (IFE). The MEC instrument discussed in this work can currently deliver up to 100 J in a 10 ns pulse. Assuming a typical scaling relation of $P \propto I^{2/3}$ (as in Eqn. 11), this would result in an ablation pressure $20 \%$ higher than achieved here. The ongoing MECU\cite{dyerMatterExtremeConditions2021} upgrade is expected to be able to deliver up to 1 kJ pulses at a similar wavelength to the MEC, and while a quantitative prediction cannot be made for such a large increase in energy (in part due to factors such as laser focusing and target preheating), this should result in a further several-fold increase in achievable pressures. Quantitative analysis of experimental images utilizing the modeling capabilities demonstrated here is underway, and will be the subject of a future work.

We have also quantified the sensitivity of simulated shock characteristics to radiation model and heat conduction flux limiter. All examined radiation schemes give similar results and reproduce empirically-observed shock velocities. A wide range of heat conduction flux limiters (0.03-0.10) produce velocities within experimental error, indicating that the system is within the realm of validity of Spitzer-H{\"a}rm heat conduction models. That Spitzer--H{\"a}rm conduction is valid in this regime is not surprising, but the level of insensitivity to flux limiter choice is unexpected. This insensitivity to both flux limiter and radiation model, combined with the small but consistent offset between simulated and empirically observed shock velocities, may indicate that there are other important computational parameters for laser-driven SBI at these pressures. However, distinguishing missing physics from experimental error will require simulation of additional experiments to increase the experimental sample size. We will expand our benchmarked dataset and examine the importance of additional physics, such as plasma viscosity, ionization and equation of state model, and non-local heat conduction in a future work. Of these, varying ionization treatment appears to have the largest potential impact on shock strength in our simulations. Plasma viscosity is also of interest due to its potential effect on the void itself, including its role in controlling the growth of instabilities. Nonetheless, these findings suggest that more complex transport models are not required to reproduce experimental results, reducing the computational expense of employing simulation-augmented image analysis.

Our findings have elucidated the roles of radiation and heat transport in HED, laser-driven SBI, and have shown that the associated instabilities in divergent geometries are driven baroclinically, as in the low pressure case. Radiation transport acts as an additional energy pathway, lowering shock velocities and pressures. Higher heat conduction increases the initial shock velocity, but this relationship rapidly inverts as the system develops. The relationship between shock pressure and ablation pressure varies with time, suggesting a non-trivial coupling between heat conduction and other physics that is more complicated than the conventional picture\cite{drakeHighEnergyDensityPhysics2018}. However, aside from a momentary spike during the shock transit of the high-impedance aluminum layer, the time-averaged shock pressure always decreases with higher heat conduction. This is due to the suppression of secondary shocks by enhanced heat conduction, which would otherwise contribute to the average pressure of the primary shock as the primary and secondary periodically coalesce. This finding indicates that, when a steady shock is desired, more conductive materials should be used, and high impedance gradients should be avoided. 


%
%

%

\begin{acknowledgments}
We wish to acknowledge the work of Yanyao Zhang, who provided the measurement of the un-shocked density of SU-8 used in this work. This work was supported by US NNSA under grant DE-NA0003914 and DE-NA0004134. Partial support from grants NSF PHY-2020249, DE-SC0020229, and DE-SC0019329 is also acknowledged. H. Aluie was also supported by US NSF grants OCE-2123496, PHY-2206380, US NASA grant 80NSSC18K0772, and US NNSA grant DE-NA0003856. Work at SLAC National Accelerator Laboratory was supported by the U.S. Department of Energy (DOE) Office of Science, Office of Fusion Energy Sciences, under the Early Career Award, 2019, for A. Gleason. Use of the Linac Coherent Light Source (LCLS), SLAC National Accelerator Laboratory, is supported by the U.S. Department of Energy, Office of Science, Office of Basic Energy Sciences under Contract No. DE-AC02-76SF00515. The experiments were performed at the Matter in Extreme Conditions (MEC) instrument of LCLS (Gleason LV08-LX49, Hart X437), supported by the DOE Office of Science, Fusion Energy Science under Contract No. SF00515. Los Alamos National Laboratory is managed by Triad National Security, LLC, for the National Nuclear Security Administration of the U.S. Department of Energy under contract 89233218CNA000001. Computing was performed on LANL supercomputing facilities.
\end{acknowledgments}

\section*{Author Declarations}

The authors have no conflicts to disclose.

\section*{Data Availability Statement}

The data that support the findings of this study are available from the corresponding author upon reasonable request.

\section*{References}
\nocite{*}

\bibliography{Refs.bib}

\end{document}